\documentclass[12pt]{article}
\usepackage{cite}
\usepackage{xcolor}
\usepackage{graphicx}
\usepackage{amsmath,amssymb,amsfonts}
\usepackage{wrapfig}
\usepackage{float}
\usepackage{geometry}
\usepackage{array}
\usepackage{multirow}
\usepackage{booktabs} 
\usepackage{indentfirst,csquotes}
\usepackage{paralist,hyperref,titlesec,fancyhdr,etoolbox}
\hypersetup{colorlinks=true, linkcolor=black, filecolor=black, urlcolor=black }
\usepackage{setspace}  
\title{GPR Full-Waveform Inversion through Adaptive Filtering of Model Parameters and Gradients Using CNN}
\author{Peng Jiang\thanks{Peng Jiang, Kun Wang, Shengjie Qiao are with the State Key Laboratory for Tunnel Engineering, the Institute of Geotechnical and Underground Engineering, the School of Qilu Transportation, Shandong University, Jinan, China (e-mail: sdujump@gmail.com, wkun2022@gmail.com, 1159003542@qq.com).}, Kun Wang, Jiaxing Wang, Zeliang Feng, Shengjie Qiao,
\\
Runhuai Deng and Fengkai Zhang
\thanks{Fengkai Zhang, Jiaxing Wang, Zeliang Feng, Runhuai Deng are with the State Key Laboratory for Tunnel Engineering, the Institute of Geotechnical and Underground Engineering, the School of Civil Engineering, Shandong University, Jinan, China (e-mail: sduzfk@163.com, 1316072258@qq.com, 15098766051@163.com, 202315038@mail.sdu.edu.cn).}}

\date{ }

\begin{document}
\renewcommand{\thefootnote}{\fnsymbol{footnote}}
\footnotetext{Corresponding author: Fengkai Zhang.}
\renewcommand{\thefootnote}{\arabic{footnote}}

\maketitle

\begin{abstract}
GPR full-waveform inversion optimizes the subsurface property model iteratively to match the entire waveform information. However, the model gradients derived from wavefield continuation often contain errors, such as ghost values and excessively large values at transmitter and receiver points. Furthermore, models updated based on these gradients frequently exhibit unclear characterization of anomalous bodies or false anomalies, making it challenging to obtain accurate inversion results. 
To address these issues, we introduced a novel full-waveform inversion (FWI) framework that incorporates an embedded convolutional neural network (CNN) to adaptively filter model parameters and gradients. Specifically, we embedded the CNN module before the forward modeling process and ensured the entire FWI process remains differentiable. This design leverages the auto-grad tool of the deep learning library, allowing model values to pass through the CNN module during forward computation and model gradients to pass through the CNN module during backpropagation. Experiments have shown that filtering the model parameters during forward computation and the model gradients during backpropagation can ultimately yield high-quality inversion results.
\end{abstract}

\section{Introduction}
\label{sec:introduction} 

Ground-Penetrating Radar (GPR) is a geophysical exploration technique that employs high-frequency electromagnetic waves to detect subsurface targets. It operates by emitting these waves into the ground through a transmitting antenna. When the waves encounter targets with varying dielectric properties, they are reflected and subsequently recorded by a receiving antenna. Researchers analyze the received radar waveforms, which include parameters such as electromagnetic field strength, amplitude, spectral characteristics and two-way travel time, to deduce the type and distribution of subsurface targets. GPR offers advantages such as high efficiency, high resolution and sensitivity to water. Unlike many acoustic and seismic waves, GPR’s electromagnetic signals propagate more quickly and do not require coupling agents, thereby expanding their range of applicability. 
Due to its high resolution, GPR can provide detailed images that aid engineers and scientists in precise subsurface analysis. 
GPR has been widely used in civil engineering site investigation~\cite{grandjean2000}, tunnel lining quality inspection~\cite{cardarelli2003} and ahead geological prospecting~\cite{zhou2007,li2017}.

For subsurface analysis, full-waveform inversion (FWI), initially proposed by Tarantola~\cite{tarantola1984} for seismic exploration, demonstrates remarkable performance in high-resolution imaging, which has drawn the attention of GPR researchers~\cite{moghaddam2005,wang2007,minet2010}. The traditional FWI approach involves minimizing the misfit between forward-modeled waveforms and actual observed waveforms and deriving gradients to update the physical parameters of the medium, such as dielectric constant and conductivity. Therefore, two main factors usually affect the inversion result: the forward modeling method and the gradient computation algorithm. For forward modeling, the finite-difference time-domain (FDTD) method~\cite{he2009} is a commonly used technique in ground-penetrating radar (GPR) due to its high precision and robustness. For gradient computation, the cross-correlation algorithm is prevalent in FWI as it can accurately reflect the subsurface structure~\cite{sun2022}.

However, gradients derived from the cross-correlation algorithm are particularly susceptible to interference. The intrinsic complexity and non-linearity of the data can introduce significant errors into these gradients, such as ghost values and excessively large values at transmitter and receiver points~\cite{zhang2012}. When these gradients are employed to update model parameters, they can further propagate errors in the forward modeling data. This progressive amplification of interference throughout the model updating process can negatively affect the accuracy of the final inversion results. 

To address this challenge, researchers have employed a range of techniques, including filtering methods~\cite{xiang2016}, regularization strategies~\cite{esser2016,du2021,li2024} and interference suppression algorithms~\cite{qin2023}. These techniques are implemented either in the design of the loss function or in the post-processing of gradients to enhance gradient accuracy, thereby improving the reliability of model parameter updates. Despite these advancements, however, these methods often face difficulties in striking an optimal balance between noise suppression, value smoothing and preserving the original gradient. This challenge frequently results in problems related to over-regularization.

In this work, we propose an adaptive filtering approach for model parameters and gradients during the full-waveform inversion (FWI) process to mitigate the progressive amplification of interference. This approach is implemented by integrating a convolutional neural network (CNN) into the FWI framework prior to the forward modeling process. Consequently, model values are first processed through the CNN module during forward computation. Additionally, we develop a differentiable forward modeling module that utilizes the automatic differentiation capabilities of deep learning libraries to update both model parameters and CNN parameters. As a result, model gradients also pass through the CNN module during backpropagation. Thus, the CNN module serves a dual role: filtering model parameters during forward computation and filtering model gradients during backpropagation, with the filter kernel (CNN parameters) adaptively learned throughout the FWI process. We denote this method as \emph{FWI\_CNN}.

Extensive experiments and mechanism studies are given to support our claims. The incorporation of the CNN module into the FWI framework results in significantly improved inversion outcomes compared to the conventional FWI method. Notably, the CNN module is optimized alongside the model parameters during the FWI process and does not depend on pre-existing training data, showcasing its excellent generalization capabilities.

\section{Related Work}
\label{sec:related work}
In Ground Penetrating Radar (GPR) inversion, there is a growing focus on learning-based methods in addition to ongoing efforts that follow conventional full-waveform inversion approaches. 

\subsection{Conventional GPR Full-Waveform Inversion}
Full-waveform inversion (FWI) methods leverage amplitude, phase and travel time information from complete radar waveforms, offering higher-resolution results. Originally introduced by Tarantola~\cite{tarantola1984} for seismic exploration, FWI has since evolved within the radar signal processing domain. In recent years, researchers have conducted numerous innovative studies to enhance the accuracy of GPR FWI. For instance, Kuroda \textit{et al.}~\cite{kuroda2005} presented FWI of EM wavefield data for imaging cross-borehole permittivity structures. Yue \textit{et al.}~\cite{yue2016} explored a Bayesian Markov-chain Monte-Carlo-based FWI method, improving the effectiveness of FWI. Zhang \textit{et al.}~\cite{zhang2024} proposed a gradient preprocessing method based on amplitude, enhancing deep energy and improving inversion outcomes in deeper regions. Liu \textit{et al.}~\cite{liu2018} derived a gradient formula for envelope waveform inversion (EWI), which effectively restores missing low-frequency information, offering better inversion capabilities for such data. Additionally, Feng \textit{et al.}~\cite{feng2019} introduced a multiscale and dual-parameter inversion method, providing reliable constraints, improved adaptability to noisy data and accurate reconstruction of subsurface dielectric properties.

However, FWI, as a linear optimization method, struggles with non-linear GPR inversion tasks. The progressive amplification of interference during gradient calculation and model updating can severely impact inversion results. Consequently, there remains considerable potential for improving GPR FWI methods.

\begin{figure*}[!t]
\centerline{\includegraphics[width=\textwidth]{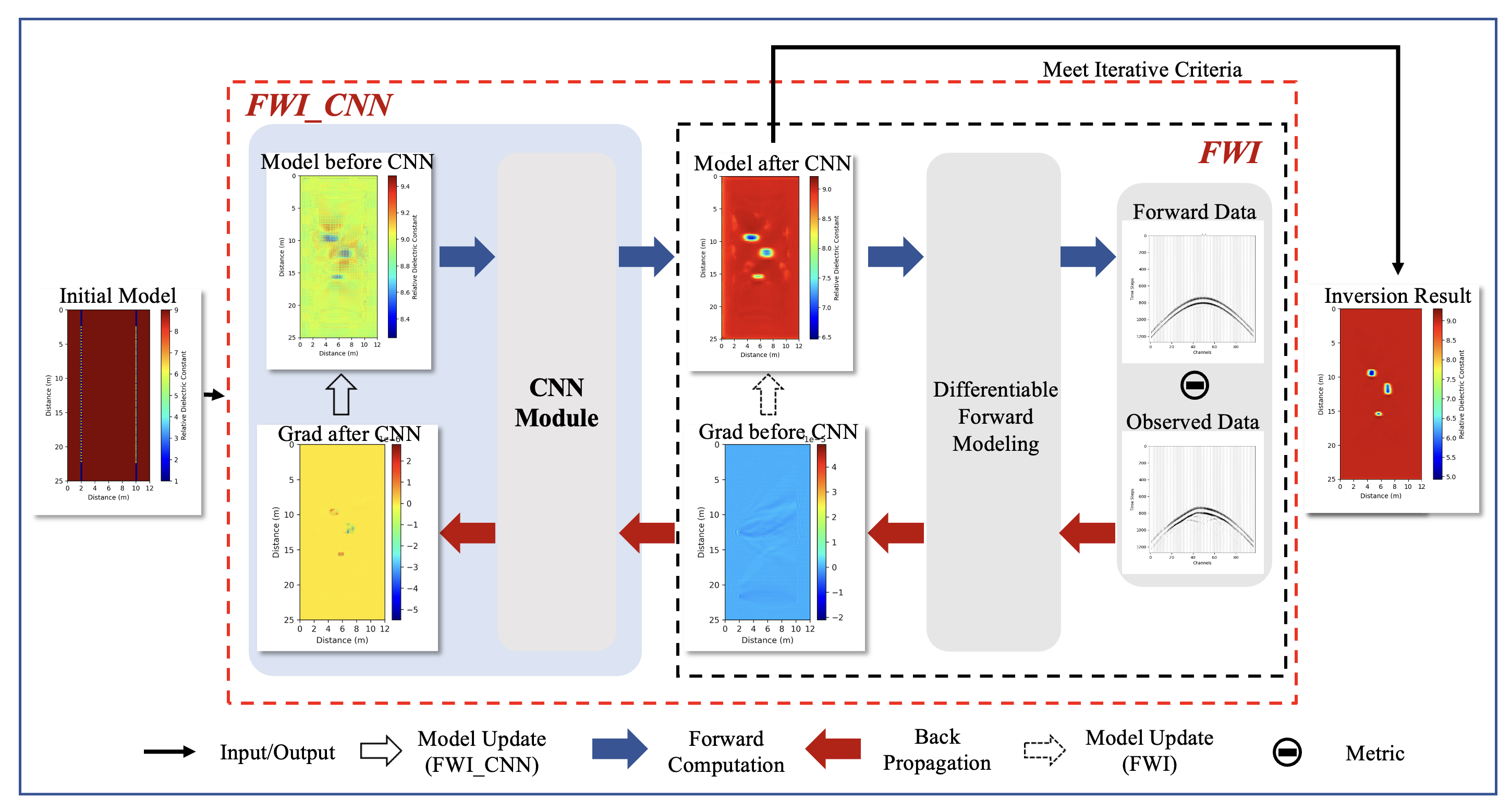}}
\caption{The proposed \emph{FWI\_CNN} framework introduces a CNN module, represented by the blue shaded area, to filter model parameters and gradients. The red dashed box outlines the main iterative process of \emph{FWI\_CNN}, while the black dashed box denotes the main iterative process of conventional \emph{FWI}. If the blue shaded area is removed from \emph{FWI\_CNN}, the remaining framework corresponds to the conventional \emph{FWI} process. The entire \emph{FWI\_CNN} framework is fully differentiable, allowing it to leverage deep learning libraries’ backpropagation tools to optimize both model parameters and CNN parameters simultaneously. The forward computation and backpropagation processes are indicated by blue and red arrows, respectively.}
\label{full}
\end{figure*}

\subsection{Deep Learning-Based GPR Inversion}
In recent years, the increasing prevalence of deep learning technologies across various computer vision tasks has highlighted the impressive capabilities of deep neural networks (DNN) in addressing complex nonlinear problems, including geophysical inversion. Alvarez \textit{et al.}~\cite{alvarez2018} demonstrated the use of a deep neural network to reconstruct underground images of concrete sewer pipes from ground-penetrating radar (GPR) B-scan images, showing that mapping GPR images to underground images using DNNs is feasible. Similarly, Liu \textit{et al.}~\cite{liu2021} introduced GPRInvNet, an encoder-decoder network designed to directly map GPR B-scan data to the relative permittivity of subsurface structures. Their results indicate that GPRInvNet effectively reconstructs complex tunnel lining defects with well-defined boundaries. 

However, the reliance on labeled datasets and the lack of real models corresponding to actual observational data present challenges for current supervised deep learning inversion frameworks, potentially leading to generalization limitations. To address these issues, Ren \textit{et al.}~\cite{ren2020} incorporated forward modeling into neural networks to achieve self-supervision between observed and synthesized data. Jiang \textit{et al.}~\cite{jiang2024} adopted a traditional FWI approach but reparameterized the model parameters using network parameters, thereby regularizing the inversion results through the network structure. This method also reduces the occurrence of local minima in high-dimensional spaces, improving the results of FWI.

Based on the aforementioned related works and insights, we observed that both conventional GPR full-waveform inversion and deep learning-based GPR inversion have their respective advantages and disadvantages. Therefore, in this work, we propose using the FWI framework as the primary structure to eliminate the dependence on large datasets while incorporating convolutional neural networks (CNN) to enhance the inversion capabilities of FWI.

\section{Method}\label{sec:method}
We first introduce our \emph{FWI\_CNN} framework in Sec.~\ref{sec:framework}, followed by a mathematical derivation in Sec.~\ref{sec:cnn} that demonstrates how the CNN module acts as a filter for model parameters and gradients. Then, in Sec.~\ref{sec:forward}, we explain the implementation of differentiable forward modeling.

\subsection{The FWI\_CNN Framework}\label{sec:framework}
As shown in Fig.~\ref{full}, the conventional FWI framework (enclosed within the black dashed box) updates the model $m_t$ using gradients $g_t$ derived from the misfit between observed data $d$ and forward modeled data $\hat{d}_t$. Starting with the initial model $m_0$ and observed data $d$, the FWI process iteratively continues until certain criteria are met, resulting in the final inversion model $m$. Instead of using the cross-correlation algorithm to derive gradients, we leverage the automatic differentiation (auto-grad) tool of the deep learning library to obtain gradients through backpropagation.

The proposed \emph{FWI\_CNN} framework (enclosed within the red dashed box) incorporates all components of the conventional FWI and adds an additional CNN module before the forward modeling. Therefore, in our framework, the model after CNN $m_t$ is the filtered result of the model before CNN $m'_t$. Similarly, the grad before CNN $g_t$ is filtered by the CNN module to obtain the grad after CNN $g'_t$. The detailed relationships between these four parameters are depicted in Fig.~\ref{full}. Unlike conventional FWI, the proposed framework updates $m'_t$ using $g'_t$. The whole process of \emph{FWI\_CNN} can be defined as:
\begin{align}
    &m_t=C_{\theta}(m'_t) \label{eq:conv},\\
    &\hat{d}_t=F(m_t),\\
    &g_t=\frac{\partial{\|d-\hat{d}_t\|}}{\partial{m_t}},\\
    &g'_t=C'_{\theta}(g_t) \label{eq:inconv},\\
    &m'_{t+1}=m'_t-\eta{g'_t},\\
    &m_{t+1}=C_{\theta}(m'_{t+1}) \label{eq:update}
\end{align}
where $\theta$ is the parameters of the CNN module. Since the CNN module acts in different ways for filtering model values in the forward computation process and model gradients in the backpropagation process, we use different notations $C$ and $C'$ to denote them, respectively. $\eta$ is the learning rate to control the update amplitude of the model value. $F$ is the forward modeling function. 

By using backpropagation to compute gradients, all parameters along the computation graph will have gradients, ensuring that the parameters of the CNN module, $\theta$, are updated during the iterative process of \emph{FWI\_CNN}. This allows the CNN module to adapt to different iterative steps without requiring any additional training data. Eq.~\eqref{eq:conv} to ~\eqref{eq:update} defines a complete iteration cycle for the proposed \emph{FWI\_CNN}. The process continues until the convergence conditions are met or the maximum number of iterations is reached. 

In \emph{FWI\_CNN}, it is essential to highlight that both the model parameters and the CNN parameters are simultaneously optimized throughout the iterative process. The final output of \emph{FWI\_CNN} is the model parameters after being processed by the CNN module, denoted as $m'_t$.

\subsection{The CNN Module as Filter}
\label{sec:cnn}

\begin{figure}[h]
\centerline{\includegraphics[width=4in]{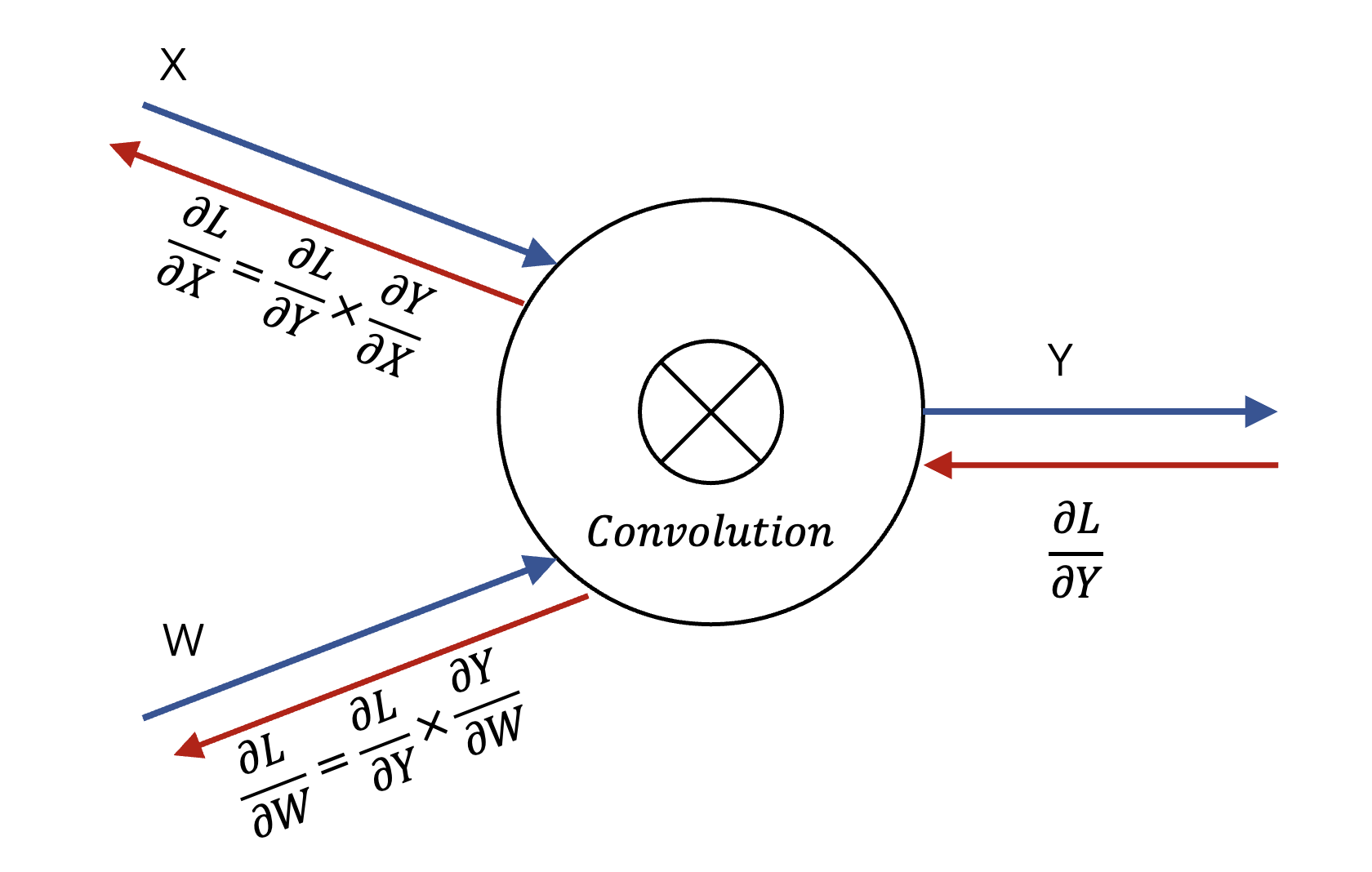}}
\caption{Computation graph for the convolution operation of a CNN layer.}
\label{conv}
\end{figure}

As we all know, a CNN consists of many convolutional layers that involve filtering with certain kernels. Therefore, Eq.~\eqref{eq:conv} can be considered as a hybrid operation to filter model parameters during forward computation. 
For simplification, we assume Eq.~\eqref{eq:conv} only has one convolutional layer. The forward pass equation for a convolutional layer is:
\begin{equation}
\begin{aligned}
 Y = X * W + b 
\end{aligned}
\label{eq:fw}
\end{equation}
where $X$ is input and $Y$ is output. We use $*$ to denote the convolution operation and $W$ to represent the convolution kernel. $b$ is the bias term. Given the convolution operation and the final loss $L$, we can define the gradients with respect to the output $Y$ as $\frac{\partial L}{\partial Y}$. The corresponding computation graph is shown in Fig.~\ref{conv}. According to the chain rule and the principles of backpropagation, we have
\begin{equation}
\begin{aligned}
     \frac{\partial L}{\partial W} &= X * \frac{\partial L}{\partial Y} ,\\
     \frac{\partial L}{\partial X} &= \frac{\partial L}{\partial Y}\frac{\partial Y}{\partial X} = \frac{\partial L}{\partial Y} * W_{\text{flip}},
\end{aligned}
\label{eq:bp}
\end{equation}
where $*$ still denotes the convolution operation and $W_{\text{flip}}$ represents the convolution kernel $W$ flipped both horizontally and vertically, essentially a transposed kernel. This equation indicates that the gradient with respect to the input $X$ is obtained by applying the flipped convolution kernel $W_{\text{flip}}$ to the gradient $\frac{\partial L}{\partial Y}$. Specifically, this process is akin to filtering the gradient $\frac{\partial L}{\partial Y}$ using $W_{\text{flip}}$. This functionality mirrors the operation performed by the CNN module in Eq.~\eqref{eq:inconv}.

In a multi-dimensional convolutional layer, the same principle applies, extended to higher dimensions and multiple channels. Consequently, the CNN module serves as a filter in both the forward computation process (as described by Eq.~\eqref{eq:conv}) and the backpropagation process (as described by Eq.~\eqref{eq:inconv}).

\subsection{Differentiable Forward Modeling}\label{sec:forward}
In this work, we use the two-dimensional TE wave FDTD formulation for forward modeling.
\begin{equation}
\begin{aligned} 
    &E_{x}^{n+1}\left(i+\frac{1}{2},j\right) = CA(m) \cdot E_{x}^{n}\left(i+\frac{1}{2},j\right) + \frac{CB(m)}{\Delta y} \cdot \\
     &\left[ H_{z}^{n+\frac{1}{2}}\left(i+\frac{1}{2},j+\frac{1}{2}\right) - H_{z}^{n+\frac{1}{2}}\left(i+\frac{1}{2},j-\frac{1}{2}\right) \right],
\end{aligned}
\label{eq:ex}
\end{equation}
\begin{equation}
\begin{aligned}
    &E_{y}^{n+1}\left(i,j+\frac{1}{2}\right) = CA(m) \cdot E_{y}^{n}(i,j+\frac{1}{2}) - \frac{CB(m)}{\Delta x} \cdot \\
     &\left[H_z^{n+\frac{1}{2}}(i+\frac{1}{2},j+\frac{1}{2}) - H_z^{n+\frac{1}{2}}(i-\frac{1}{2},j+\frac{1}{2})\right],
\end{aligned}
\label{eq:ey}
\end{equation}
\begin{equation}
\begin{aligned}
&H_{z}^{n+\frac{1}{2}}\left(i+\frac{1}{2},j+\frac{1}{2}\right) = CP(m) \cdot H_{z}^{n-\frac{1}{2}}\left(i+\frac{1}{2},j+\frac{1}{2}\right) \\
&- \left\{ \frac{CQ(m)}{\Delta x} \cdot \left[ E_y^n\left(i+1,j+\frac{1}{2}\right) - E_y^n\left(i,j+\frac{1}{2}\right) \right] \right. \\
&- \left. \frac{CQ(m)}{\Delta y} \cdot \left[E_{x}^{n}\left(i+\frac{1}{2},j+1\right) - E_{x}^{n}\left(i+\frac{1}{2},j\right) \right] \right\}.
\end{aligned}
\label{eq:hz}
\end{equation}

Here, $E$ represents the electric field strength, and $H$ represents the magnetic field strength. The superscript in the upper right corner indicates sampling with respect to the time axis $t$. 
The value of $m$ corresponds to the spatial position of the field component nodes on the left side of the two-dimensional TE wave formula. Once the medium parameters and time parameters for forward modeling are determined, the values of CA, CB, CP, and CQ are fixed, which do not change with time iterations. 

During each iteration, the electric field components $E_x$ and $E_y$ are updated based on the spatial derivatives of the magnetic field component $H_z$, while the magnetic field component $H_z$ is updated based on the spatial derivatives of the electric field components $E_x$ and $E_y$. The spatial derivatives can be represented using convolution operations. For instance, consider the computation of Eq.~\eqref{eq:ex}, its spatial derivative part can be reformulated as,

\begin{figure*}[t]
\centerline{\includegraphics[width=\textwidth]{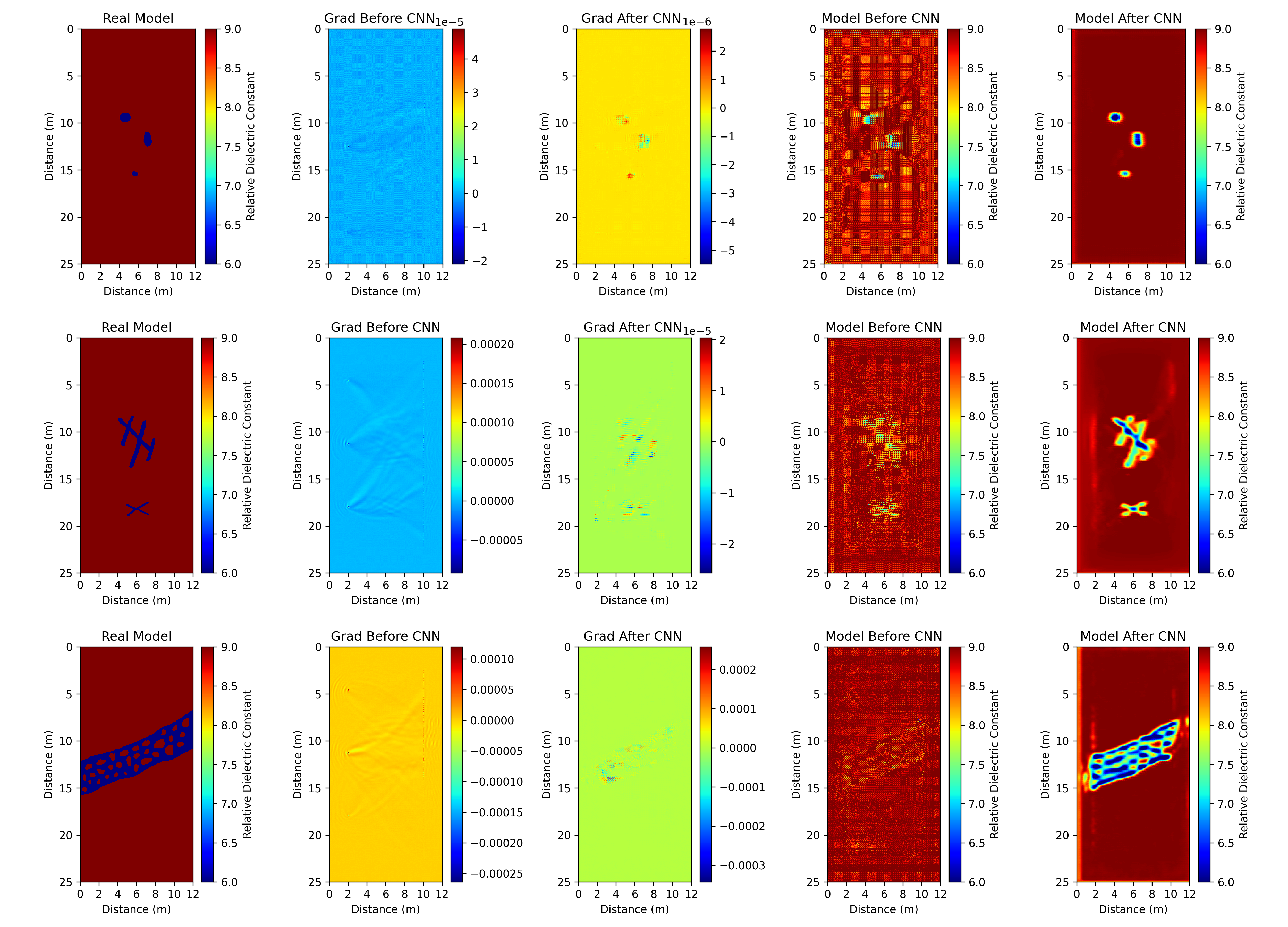}}
\caption{Comparison of the filter effects for model parameters and gradients by the CNN module. From top to bottom are the Karst cave model, the fracture model and the fault model. From left to right are the real model, grad before CNN, grad after CNN, model before CNN and model after CNN. To facilitate comparison with the real model, post-processing steps such as clipping and scaling were applied to both the model before CNN and the model after CNN to standardize the color maps.}
\label{grad}
\end{figure*}

\begin{equation}
\begin{aligned}
    &H_z^{n+\frac12}(i+\frac12,j+\frac12)-H_z^{n+\frac12}(i+\frac12,j-\frac12)\\
    &=\left[H_z^{n+\frac12}(i+\frac12,j-\frac12),H_z^{n+\frac12}(i+\frac12,j+\frac12)\right]^T * W_y,
\end{aligned}
\label{eq:hz_diffy}
\end{equation}
where $W_y$ represents the convolution kernel with kernel values $[-1, 1]^T$, $*$ represents the convolution operation.
Therefore, the entire electric field components $E_x$ and $E_y$, as well as the magnetic field component $H_z$, can be expressed as:
\begin{equation}
\begin{aligned}
    &E_{x}^{n+1} = CA\cdot E_{x}^{n} +\frac{CB}{\Delta y}\cdot \left( H_z^{n+\frac{1}{2}} * W_y \right),
\end{aligned}
\label{eq:ex_conv}
\end{equation}
\begin{equation}
\begin{aligned}
    &E_{y}^{n+1} = CA\cdot E_{y}^{n} -\frac{CB}{\Delta x}\cdot \left( H_z^{n+\frac{1}{2}} * W_x \right),
\end{aligned}
\label{eq:ey_conv}
\end{equation}
\begin{equation}
\begin{aligned}
    H_{z}^{n+\frac{1}{2}} = CP \cdot H_{z}^{n-\frac{1}{2}} & - \left[ \frac{CQ}{\Delta x} \cdot \left( E_y^n * W_x \right) \right.  \quad \\
    & - \left. \frac{CQ}{\Delta y} \cdot \left( E_{x}^{n} * W_y \right) \right].
\end{aligned}
\label{eq:hz_conv}
\end{equation}

Here, $W_x = [-1, 1]$ represents the convolution kernel in the x-direction, while $W_y = [-1, 1]^T$ is the convolution kernel in the y-direction. Consequently, the field values $E_x$, $E_y$, and $H_z$ can be expressed in convolutional form, allowing for implementation with a deep learning library to ensure differentiability.

\subsection{Loss Definition}
Typically, conventional full-waveform inversion (FWI) measures the misfit between observed data $d$ and forward modeled data $\hat{d}_t$ using the $L_2$ norm. However, in our \emph{FWI\_CNN} framework, we introduce a CNN module with randomly initialized parameters. This often results in the model value $m_t$ containing significant noise. To mitigate this issue, we apply a regularization term to smooth $m_t$ during the iterations. Consequently, the loss function for the $t$-th iteration in our framework is defined as:
\begin{equation}
L_t=\|\hat{d}_t-d\|+\lambda\mathcal{R}(m_t)
\label{eq1}
\end{equation}
where $R(m)$ is the regularization term. Specifically, we use Total Variation (TV) as a regularization term in this work, and $\lambda$ is the weight used to balance the strength of regularization. 

\begin{figure*}[t]
\centerline{\includegraphics[width=\textwidth]{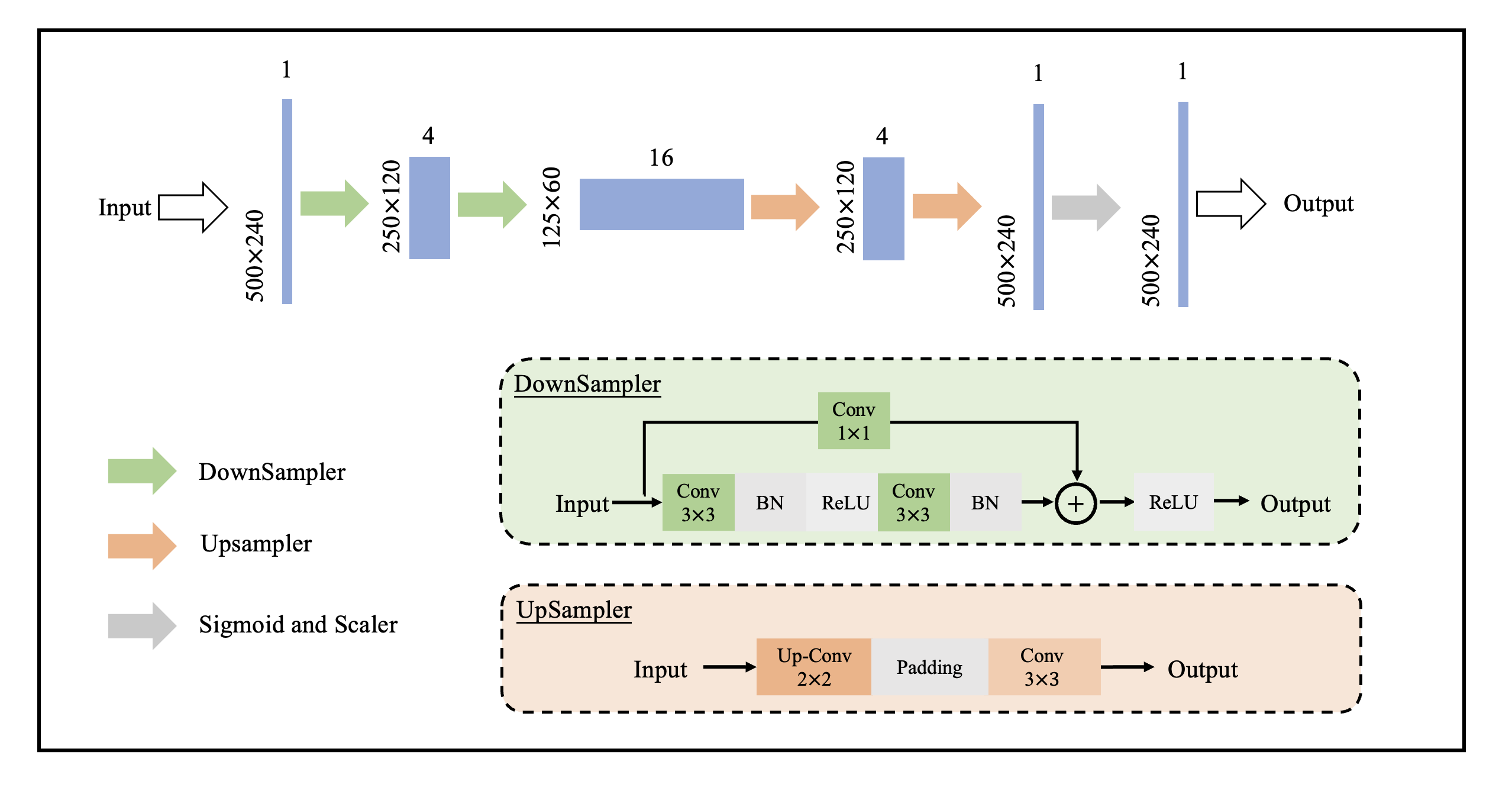}}
\caption{The structure of CNN module. The network consists of four modules: two downsampling modules and two upsampling modules, ultimately producing an output $m'_k$ with the same shape as the input model parameters $m_k$.}
\label{net}
\end{figure*}

\section{Experiments}
\subsection{Experiment Data Preparation}
To verify our method, we designed three two-dimensional models: the Karst Cave model, the Fracture model and the Fault model. Each model is of size $25m \times 12m$ with a grid size of 0.05m, resulting model with $120,000$ parameters. In the experiment, only the relative permittivity of the model is used for single-parameter inversion. The background relative permittivity is set to 9, and the relative permittivity of the model anomaly is set to 6. To minimize boundary reflections, we employ a sponge-absorbing boundary condition. This involves adding sponge layers around the original model. 

\begin{table}[t]
\centering
\caption{Network structure details.}
\label{table:net}
{
\begin{tabular}{l l c c l} 
\toprule
& Type & Filter & Stride & Output Size \\
\midrule
Input & - & - & - & \begin{tabular}[c]{@{}l@{}}(500,240,1)\\ $H \times W \times C$ \end{tabular} \\
\midrule
\multirow{3}{*}{DownSampler\_1} & Conv & $3 \times 3$ & 2 & (250,120,4) \\
 & Conv & $3 \times 3$ & 1 & (250,120,4) \\
 & Conv & $1 \times 1$ & 2 & (250,120,4) \\
\midrule
\multirow{3}{*}{DownSampler\_2} & Conv & $3 \times 3$ & 2 & (125,60,16) \\
 & Conv & $3 \times 3$ & 1 & (125,60,16) \\
 & Conv & $1 \times 1$ & 2 & (125,60,16) \\
\midrule
\multirow{2}{*}{UpSampler\_1} & Up-conv & $2 \times 2$ & 2 & (250,120,4) \\
 & Conv & $3 \times 3$ & 1 & (250,120,4) \\
\midrule
\multirow{2}{*}{UpSampler\_2} & Up-conv & $2 \times 2$ & 2 & (500,240,1) \\
 & Conv & $3 \times 3$ & 1 & (500,240,1) \\
\midrule
\multirow{2}{*}{PostProcess} & Sigmoid & - & - & (500,240,1) \\
 & Scaler & - & - & (500,240,1) \\
\bottomrule
\end{tabular}
}
\end{table}

We use the differentiable forward modeling described in Sec.~\ref{sec:forward} and a cross-hole radar observation setup to obtain the observed data. The detailed observation setup is illustrated in Initial Model of Fig.~\ref{full}. In this setup, the transmitter point borehole is positioned at a horizontal distance of 2m, and the receiver point borehole is positioned at a horizontal distance of 10m. The horizontal distance between the two boreholes is 8 meters, and both boreholes have a depth of 25 meters. The transmitter and receiver points begin 2.5 meters underground, comprising 48 transmitters and 96 receivers. To satisfy the numerical stability conditions, we set $nt$ to 1271 in the above experimental setup, resulting in a total observed data dimension of $48 \times 96 \times 1271$.

\subsection{Experiment Hyperparameter Setting}
\label{sec:hyperparameter}
We compare our framework, \emph{FWI\_CNN}, with conventional Full-Waveform Inversion (\emph{FWI}) and FWI with Total Variation regularization (\emph{FWI\_TV}). All methods share similar components, with the primary distinction being that \emph{FWI\_CNN} incorporates an additional CNN module. For these methods, we perform multiscale inversion using Ricker wavelets with central frequencies of 60 MHz, 80 MHz, and 100 MHz. Each frequency is processed for 100 iterations, leading to a total of 300 iterations. Additionally, these methods use blank initial models for inversion.

We use the Adam optimizer with an initial learning rate of 0.01 to iteratively update the model parameters $m_t$ and the CNN module parameters $\theta$. Random sampling is employed among the 48 transmitter points, with a batch size of 3, meaning three random transmitter points are selected for each inversion step. The initial learning rate is gradually decreased during the iterations to ensure convergence.

The experiments were conducted using multiple NVIDIA A100 SXM GPUs, with Distributed Data Parallel (DDP) employed for parallel computing.

\begin{figure*}[t]
\centerline{\includegraphics[width=\textwidth]{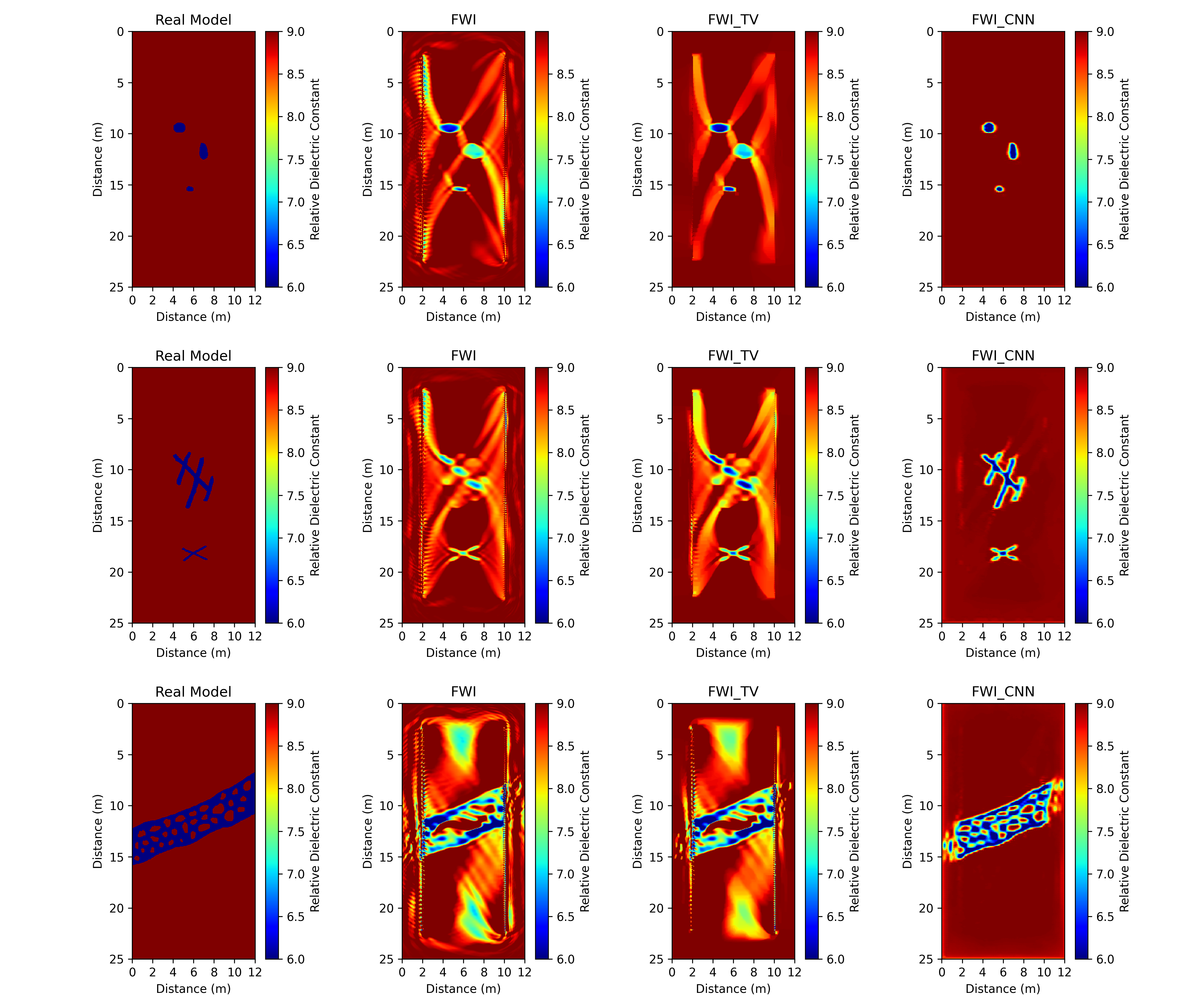}}
\caption{The visual comparison of inversion results is shown as follows: from left to right are the real model, the inversion results for conventional \emph{FWI}, \emph{FWI\_TV} and \emph{FWI\_CNN}. From top to bottom are the Karst cave model,
the fracture model and the fault model. Consistent with Fig.~\ref{grad}, post-processing operations, including clipping and scaling, were applied to the results to standardize the color maps for better comparison with the real model.}
\label{fig:results}
\end{figure*}

\subsection{CNN Module Structure}
The CNN module is illustrated in Fig.~\ref{net}. It is fundamentally a bottleneck neural network, comprising two downsampler modules, two upsampler modules and a postprocess module. Each downsampler module is implemented using residual blocks, while the upsampler modules use simple deconvolution operations for upsampling.
Generally, the downsampler compresses the spatial dimensions of the input and extracts additional feature maps to encode information, while the upsampler progressively restores the spatial and channel dimensions to decode information. Tab.~\ref{table:net} details the settings for each layer of the CNN module used in this work. Notably, the total number of parameters in the CNN module is only $3,787$, which constitutes just 3\% of the total parameters optimized throughout the entire process.

\begin{figure*}[t]
\centerline{\includegraphics[width=\textwidth]{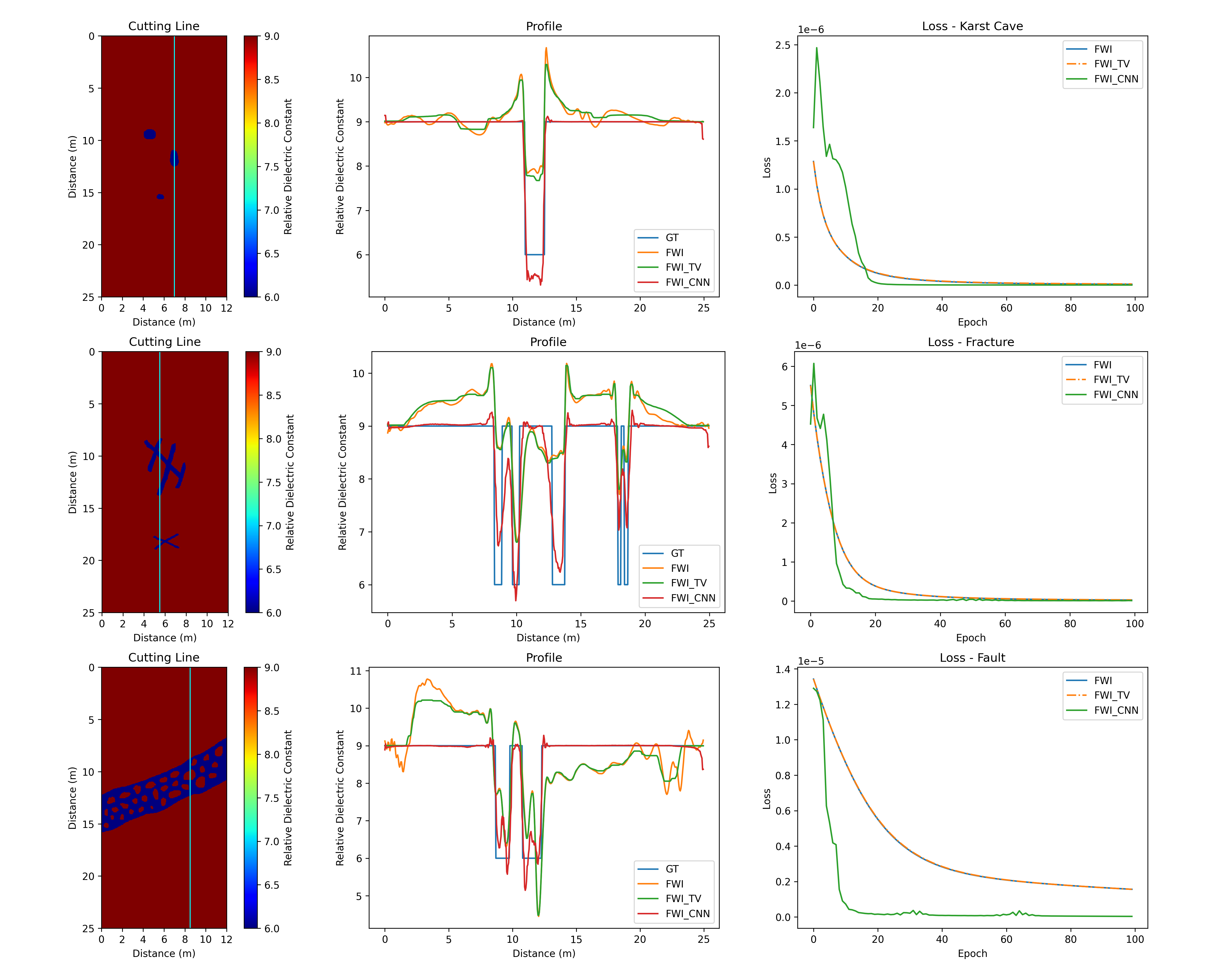}}
\caption{The comparison includes profiles and loss curves for three models, arranged as follows: from top to bottom are the Karst cave model, the fracture model, and the fault model. A vertical cutting line was randomly selected from each experimental model. From left to right, the figures show the positions of the chosen cutting lines, the profile comparisons at these cutting line positions, and the loss curves. For simplicity, only the loss curve for one scale is shown.}
\label{baseline}
\end{figure*}

\subsection{Mechanism Study}
First, we conduct experiments to validate our claim that embedding the CNN module in the FWI framework allows it to function effectively as a filter during both the forward computation and backpropagation stages. In Fig.~\ref{grad}, we illustrate the morphology of model parameters and gradients before and after processing through the CNN module. For clarity, only the results of one iteration are displayed.

In the ``Grad Before CNN'' images, we observe excessively large values and ghost values at transmitters. These abnormalities are indicative of issues in gradient calculation. After applying the CNN module during backpropagation, as shown in the ``Grad After CNN'' images, the abnormal gradient values are effectively suppressed, and more accurate gradients are obtained. However, some residual noise remains in these gradients, which can still affect the subsequent inversion process. The ``Model Before CNN'' images reveal that the noisy gradients lead to a model with significant noise. However, after processing through the CNN module during the forward computation, as shown in the ``Model After CNN'' images, the model is significantly refined. 

The experimental results confirm our claims. Notably, the CNN module parameters are optimized concurrently with the model parameters during the optimization process. This concurrent optimization ensures that the CNN module adaptively adjusts the parameters to filter both model parameters and gradients, enhancing the overall performance of the inversion.

In ``Grad Before CNN'', ``Grad After CNN'' and ``Model Before CNN'', checkerboard artifacts are noticeable in these intermediate results. These artifacts are inevitably caused by the use of deconvolution for upsampling. Replacing deconvolution with interpolation for upsampling can mitigate this effect~\cite{odena2016}. In this work, we use deconvolution for upsampling and incorporate TV regularization into the loss function, effectively smoothing out the checkerboard artifacts.

\subsection{Comparison Results}
In this section, we compare the proposed \emph{FWI\_CNN} with conventional \emph{FWI} and \emph{FWI\_TV}, as detailed in Sec.\ref{sec:hyperparameter}. The visual results are presented in Fig.~\ref{fig:results}. It is evident that \emph{FWI\_CNN} excels in characterizing anomaly shapes, approximating numerical values, and suppressing interference, particularly the ghosting values caused by progressive amplification during optimization. Additionally, our method effectively mitigates excessively large values at transmitter and receiver points. In the first model, since the anomalies are relatively simple, the artifact values and the values at the transmitter and receiver points are well suppressed. In the second and third models, due to the complexity of the models, some interference values still appear, but our method still shows significant advantages compared to traditional methods. These results demonstrate the robustness and accuracy of \emph{FWI\_CNN} in complex geological conditions, enhancing its capability to detect and locate anomalies. 

By comparing \emph{FWI} with \emph{FWI\_TV}, it is evident that traditional regularization methods, such as the Total Variation (TV) regularizer, can help achieve smoother inversion results. However, they do not fully eliminate errors, as they lack the adaptive capability to address varying types of errors effectively. We further present the permittivity profiles and loss curves in Fig.~\ref{baseline}, which highlight the superiority of our \emph{FWI\_CNN} in both numerical approximation and convergence efficiency.

\begin{table}[h]
    \centering
    \renewcommand{\arraystretch}{1.5}
    \caption{Quantitative Comparison}
    {
        \begin{tabular}{c c c c c}
            \toprule
            \multirow{2}{*}{\textbf{Model}} & \multirow{2}{*}{\textbf{Metric}} & \textbf{FWI} & \textbf{FWI\_TV} & \textbf{FWI\_CNN} \\ 
            & & \textbf{Results} & \textbf{Results} & \textbf{Results} \\
            \midrule
            \multirow{4}{*}{\textbf{Karst Cave}} & MAE $\downarrow$  & 0.18701 & 0.13815 & \textcolor{red}{0.01714} \\ 
            & MSE $\downarrow$  & 0.09373 & 0.06374 & \textcolor{red}{0.00926} \\ 
            & SSIM $\uparrow$ & 0.63324 & 0.84658 & \textcolor{red}{0.97154} \\ 
            & MSSIM $\uparrow$ & 0.77708 & 0.93766 & \textcolor{red}{0.97898} \\ 
            \midrule
            \multirow{4}{*}{\textbf{Fracture}} & MAE $\downarrow$  & 0.77982 & 0.66817 & \textcolor{red}{0.19984} \\ 
            & MSE $\downarrow$  & 1.39113 & 1.21572 & \textcolor{red}{0.29816} \\ 
            & SSIM $\uparrow$ & 0.28974 & 0.49870 & \textcolor{red}{0.84073} \\ 
            & MSSIM $\uparrow$ & 0.46972 & 0.68426 & \textcolor{red}{0.86121} \\ 
            \midrule
            \multirow{4}{*}{\textbf{Fault}} & MAE $\downarrow$  & 0.31451 & 0.27413 & \textcolor{red}{0.08570} \\ 
            & MSE $\downarrow$  & 0.25548 & 0.21999 & \textcolor{red}{0.05811} \\ 
            & SSIM $\uparrow$ & 0.50629 & 0.70819 & \textcolor{red}{0.90046} \\ 
            & MSSIM $\uparrow$ & 0.70032 & 0.84929 & \textcolor{red}{0.92205} \\ 
            \bottomrule
        \end{tabular}
    }
    \label{tab:results}
\end{table}

Finally, we conduct a quantitative evaluation of all methods using several metrics: Mean Absolute Error (MAE), Mean Squared Error (MSE), Structural Similarity Index (SSIM) and Multi-Scale Structural Similarity Index (MSSIM). MAE and MSE assess value accuracy, while SSIM and MSSIM measure structural similarity. These metrics enable us to evaluate inversion results in terms of error magnitude and local structure fitting. The scores for all metrics are detailed in Tab.~\ref{tab:results}. As shown, our method demonstrates significant advantages across all four metrics for the three models tested. Specifically, it outperforms the other methods in all evaluation indicators, highlighting its effectiveness in enhancing inversion accuracy. 

\section{Conclusion}
GPR Full-Waveform Inversion (FWI) often produces gradients with excessively large values and ghost values, which can degrade inversion results. To address these issues, we introduce a novel FWI framework incorporating an embedded Convolutional Neural Network (CNN) module. By leveraging the auto-grad tool of a deep learning library and implementing differentiable forward modeling, both model parameters and CNN module parameters are optimized simultaneously. This enables the CNN module to adaptively adjust its parameters to filter model parameters during forward computation and model gradients during backpropagation. Comprehensive studies and experiments validate our approach and demonstrate its superiority over traditional methods. Our findings show that the proposed method effectively integrates the strengths of deep learning-based inversion techniques with conventional FWI approaches, offering promising nonlinear mapping capabilities and strong generalization without requiring labeled training data.

\end{document}